\documentclass[preprint,12pt]{elsarticle}

\usepackage{graphicx}

\usepackage{amssymb}
\usepackage{url}
\usepackage{clrscode}
\usepackage{amsthm}

\newtheorem{mythe}{Theorem}
\newtheorem{mydef}{Definition}

\journal{Knowledge-Based System Journal}

\begin{document}

\begin{frontmatter}



\title{Parallelization in Extracting Fresh Information from Online Social Network}


\author{Rui Guo}
\author{Hongzhi Wang}
\author{Mengwen Chen}
\author{Jianzhong Li}
\author{Hong Gao}

\address{Harbin Institute of Technology}

\begin{abstract}
Online Social Network (OSN) is one of the most hottest services in the past years.
It preserves the life of users and provides great potential for journalists, sociologists and business analysts.
Crawling data from social network is a basic step for social network information analysis and processing.
As the network becomes huge and information on the network updates faster than web pages, crawling is more difficult because of the limitations of band-width, politeness etiquette and computation power.
To extract fresh information from social network efficiently and effectively, this paper presents a novel crawling method and discusses parallelization architecture of social network.
To discover the feature of social network, we gather data from real social network, analyze them and build a model to describe the discipline of users' behavior.
With the modeled behavior, we propose methods to predict users' behavior.
According to the prediction, we schedule our crawler more reasonably and extract more fresh information with parallelization technologies.
Experimental results demonstrate that our strategies could obtain information from OSN efficiently and effectively.

\end{abstract}


\end{frontmatter}


\section{Introduction}
\subsection{Motivation}

Social Network Service is one of the hottest services in the last few years.
It has a tremendous group of users.
For instance, Facebook has 874 million active users \cite{facebook-cite} and Twitter reaches 500 million users.
It is estimated that at least 2.3 billion tweets have been published on Twitter during a 7-month period, more than 300 million tweets per month \cite{snap-cite}.
And The Yahoo! Firehouse has reached 750K ratings per day, 150K comments per day \cite{leskovec2011social}.

There are a few social media datasets such as Spinn3r \cite{Spinn3r-cite}.
About 30 million articles (50GB of data), including 20,000 news sources and millions of blogs were added to Spinne3r per day \cite{snap-cite}.
As people access OSN frequently, advertisement could be broadcast according to users' behavior.
\cite{byun2012automated} studies various Super Bowl ads by applying data mining techniques through Twitter messages.
Using Twitter, \cite{sakaki2010earthquake} detects earthquakes and \cite{aramaki2011twitter} studies influenza spread.

For crawling, one of the most important factors is the freshness.
Denev, Mazeika, Spaniol and Weikum \cite{denev2009sharc} designed a web crawling framework named SHARC.
It pays more attention to the relationship between freshness and time period.
Moreover, Olston and Pandey proposed a crawling strategy optimized for freshness \cite{olston2008recrawl}, it concerns more about the current time point.
Even though OSN crawling is related to web crawling, crawling OSN for fresh information is different from web crawling in following points and brings new technical challenges.

1. New messages are published more frequently.
Everyone can conveniently register, post, comment and reproduce messages at twitter.com while a server and special skill is required to maintain a website.
As a result, the hottest topic may change in few hours on the OSN.
With this feature, freshness metrics for web crawlers could not be applied on OSN crawlers perfectly.

2. The messages on OSN are shorter than web pages.
The former are often made up by a few sentences (e.g. Twitter limits its messages to 140 characters) while the latter often contain titles and long contents.

3. The OSN is closely related to users' daily life.
Ordinary people post messages in the day and not at night.
Hence we must consider people's work and rest time when crawling for OSN messages.
As a comparison, this feature is not considered in web crawling.

4. OSN network is more complex.
The relationships between close users are shown explicitly.
We can trace the friend relationship and forwarded message easier than on the web.
The web pages are usually open to everyone while some OSN mes-sages are only available for friends, such as Facebook messages.

With these features, new techniques for crawling fresh OSN information are in demand.
What's more, in a real crawler system, many machines are running the crawler program at the same time, leading to a series of problem such as task schedule and workload balance.
So it is necessary to design a parallelization crawler architecture.

With the consideration that the goal of crawling OSN information is to gather new information, this paper aims to crawl as more new messages as possible with the limited resources.

\subsection{Contributions}

For crawling, the limitations on resources include bandwidth, computation power and politeness etiquette.
For instance, for twitter.com, we may be permitted to get at most 200 tweets with a Twitter API call.
The restriction on the method call is 350 calls per hour for one authorized developer account \cite{Twitter-cite}.
Actually, the API restriction is the bottleneck of most OSN crawlers.

To leverage the limited resources and freshness requirements of the crawler, we classify the users according to their behaviors and model their behavior of updating posts respectively.
With these models, the time of post updating for different users is predicted and the crawler could access the posts of users only when corresponding is updated.
As a result, the latest information is collected with limited resources.

Combing the steps discussed above, we propose \underline{C}rawl based on \underline{U}ser \underline{Vi}siting \underline{M}odel (CUVIM in brief) based on the observations and classifications of users' behaviors.
In this paper, we focus on the messages in OSN.
Considering the relationship updating between users as a special type of message, the relationship of updating information could also be crawled with the techniques in this paper.

According to different behaviors on updating the posts, we classify OSN users into 4 kinds: the inactive account, frequently changing account, reasonable constant ac-count, and authority account.
This is the first contribution of this paper.

We design different updating time predication model and accordingly develop efficient crawling strategies.
This is the second contribution of this paper.

Concretely, for the inactive accounts and frequently changing accounts which change not very frequently, the changes can be described by the Poisson process and the changing rate can be predicted by statistics methods.
Thus we build the Pois-son model and take web crawling strategy to crawl OSN data.

From reasonable constant accounts and authority accounts who post messages frequently, we observe that the frequency of new messages is related to users' daily life.
According to this observation, we can crawl many fresh and useful messages in the day while almost no new messages at night.
And we build the Hash model to visit those active users to crawl the information efficiently.

As the third contribution, extensive experiments are performed to verify the effectiveness and efficiency of the techniques in this paper.
We crawled the last 2,000 messages of 88,799 randomly selected users.
The result shows that 80,616,076 messages are collected.
From experimental results, the Poisson process model collects 12.14\% more messages than a Round-and-Robin (RR) style method.
The Hash model collects about 50\% more messages than the RR style method.
The parallelization method limits the workload difference of the Poisson process model to less than 13.27\% of a random method.
We also tested our parallelization crawler architectures.
The results show that the speed up of the architectures are linear while the workload difference between the machines is almost ignorable.

This paper is organized as follows.
Section 2 reviews related work at crawler field;
Section 3 introduces our work and discovery from the data;
Section 4 introduces our crawl altimeters;
Section 5 introduces the parallelization technology in the crawler system;
Section 6 shows the experimental results;
Section 7 draws the conclusions and proposes further research problems.

\section{Related Work}
Only a few methods are proposed to crawl OSN data.
\cite{byun2012automated} describes a Twitter Crawler developed by Java.
They pay more attention to the implementation detail of the crawler and the data analysis.
Instead, we focus on the crawling method and develop algorithms to gather more information of the specific OSN users.

TwitterEcho is an open source Twitter crawler developed by Boanjak et al. \cite{boanjak2012twitterecho}.
It applied a centralized distributed architecture.
Cloud computing is also used for OSN crawler \cite{noordhuis2010mining}. Noordihuis et al. collect Twitter data and rank Twitter users through the PageRank algorithm.
Another attempt is to crawl parallel.
Duen et al. implemented a parallel eBay crawler in Java and visited 11,716,588 users in 23 days \cite{chau2007parallel}.
The three methods aim to get more calculating resource, while we focus on a more reasonable crawling sequence with the given resource.

Whitelist accounts were once available on Twitter.
Kwak et al. crawled the entire Twitter site successfully, including 41.7 million user profiles and 106 million tweets by Twitter API \cite{dziczkowski2009social}.
However, whitelist accounts are no longer available now. It is the same for \cite{noordhuis2010mining}.
As the API has rate-limiting now, we propose algorithms to improve the crawl efficiency.

Another related work to OSN message collector is web crawlers.
Generally the page changing follows the Poisson process model.
The Poisson process model assumes that the changing rate $\lambda_i$ in every same time unit $\Delta$ is the same.
And the changes are modeled as following formula.

\begin{displaymath}
P[\textrm{changes in $\Delta$ is $k$}]= \frac{e^{-{\lambda_i \Delta}}(\Delta \lambda_i)^k}{k!}
\end{displaymath}

This equation can predict the possible changes of web pages and can be applied on only the inactive OSN users, since they usually change constantly.
However, when considering the active OSN users who post messages frequently, the change rate is not equal all the time.
The discipline of OSN messages follows the users' daily life.
A user may post many messages in the day and much less at night, so the changing rate is not same for the day and night.

Many web crawlers are proposed.
Reprehensive measures for web crawling are sharpness \cite{denev2009sharc} and freshness \cite{olston2008recrawl}.
The strategies define sharpness or freshness to the crawling, and schedule crawling to achieve those targets.
Differently, we choose the total number of new OSN messages as our target, and schedule according to the OSN users' behavior.

There are other matured web crawling strategies.
J. Cho, H. Garcia-Molina and L. Page improve the crawling efficiency through URL ordering \cite{cho1998efficient}.
They define several importance metrics including ordering schemes and performance evaluation measures to obtain more ¡°important¡± URL first.
J. Cho, H. Garcia-Molina also pro-poses a strategy for estimating the change frequency of pages to make the web crawl-er work better \cite{cho2003estimating} by identifying scenarios and then develop ¡°frequency estimators¡±.
C. Castillo, M. Marin, A. Rodriguez and R. Baeza-Yates combine the breadth-first ordering with the largest sites first to crawl pages fast and simply \cite{castillo2004scheduling}.
J. Cho and U. Schonfeld make PageRank coverage guarantee high personalized to improve the crawler \cite{cho2007rankmass}.

\section{OSN Data Analysis And Classification}

At first, to design the proper crawler for OSN, we discuss the features of OSN data and the classification of OSN users by their behavior in this section.
Considering the four features of the OSN in Section 1, we propose novel methods.
At first, we define audience and channel as following in OSN relationships to study the OSN, where both A and B are users.

\begin{mydef}
(audience)
Audience is a one way relationship on OSN. $A$ is $B$'s audience, that means, $A$ can check $B$'s OSN messages.
\end{mydef}

\begin{mydef}
(channel)
Channel is a one way OSN relationship on OSN. $A$ is $B$'s channel, that means, $A$'s messages will be checked by $B$.
\end{mydef}

To study the behavior of OSN users of messages updating, we crawl and study OSN messages.
For effective crawling, we choose several top OSN users from the influence ranking list in Sina weibo (http://weibo.com), which is a famous OSN with millions of users.
They are added to the crawling list as the seeds and some of their channels are randomly selected.
The channels behave more active than the audiences, thus we can avoid invalid accounts in the list.
The channels are added to the crawling list.
Then they are treated as the new seeds and their channels are accessed.
We can end the iterations when we get enough users in the crawling list.

With $a$ seeds, $k$ channels are chosen and $n$-hop channels for each seed are traversed, the crawling list contains $\sum \matrix{
  n \hfill \cr
  i = 0 \hfill \cr}  a{k^n} = a({k^{n + 1}} - 1)/(k - 1)$ users.
From this formula, the smaller $n$ is, the more representative the users in the list are: at the beginning the seed users are famous, and later more and more ordinary people are added into the list.
Since without any information, it is impossible to predict the frequency of posting, initially, we have crawled the data for all users in a Round-and-Robin style.

We crawled 10,000 users' data for two months, and got 1,853,085 messages in total.
According to the experimental results, we have following observations.

1. Users are quite different in posting frequency.
The users in the crawling list have at least one audience.
It means that those accounts are valid and are or were once active.
However, during the experiment period, about 1/2 users post less than one message per day.
And 1/5 users post more than 10 messages per day.
This observation means that the crawl frequency for different users should be different.

2. The frequency of new messages may change with time.
An extreme user post 18 messages in the first day but did not post anything in the following three days.

3. The frequency of posting new messages is related to the users' daily life.
Experimental result shows that an actress posts about half messages in the late night while a manager posts most messages in the afternoon.

4. Some accounts are maintained by professional clerks or robots, such as a newspaper's official account.
Those accounts have more audiences, change more frequently and have more influence than the personal users.

With above observations, we classify all users into 4 types by their behavior.
To illustrate the features of these four kinds of users, we show the experimental results of four user's belonging to each type in Figure 1 to 4.
It shows the total number of new messages in each 15 minutes of a day.
The horizontal axis means time from 00:00 to 24:00 in a day, and the longitudinal axis means the total number of new messages post with 15 minutes as the unit.
In Figure 1, 3 and 4, the line in the figure means the number of all messages in the 2 months.
In Figure 2, each line means a day.

\begin{figure}
  \centering
  \includegraphics[width=0.8\textwidth]{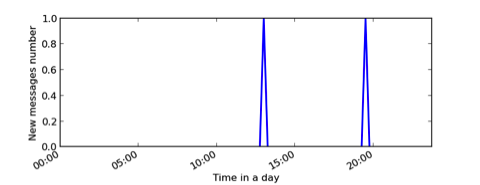}
  \caption{The Updating Ratio of an Inactive Account}
\end{figure}

1. Inactive account.
The users in this type post nothing in a long period or have little channels and audiences.
From Figure 1, it can be observed that the figure of inactive account has at most a few points.
It means that the user has hardly post any messages in a day, but may a few messages in a few weeks.
When observing for a long period, those users behave as web pages.
It means that an inactive user may post three messages a month while a page may change three times this month.
The number of possible changes between each equal time unity $\Delta$ when $\Delta$ is large enough.
Thus we can describe the behavior of this type by Poisson Process.

\begin{figure}
  \centering
  \includegraphics[width=0.8\textwidth]{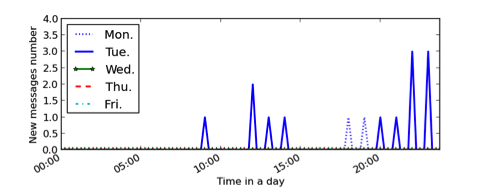}
  \caption{The Updating Ratio of an Instable Changing Account}
\end{figure}

2. Instable changing account.
Such type of users' behavior is instable and cannot be predicted.
Users in this class are often very active at a short period and remain silence later.
For example, the one who post 18 messages and did not post anything in later 3 days.
It is hard to design an effective crawling strategy for those people.

Figure 2 shows the behavior of a instable changing account.
The user posted 2 messages on Monday, 13 messages on Tuesday and nothing in the later three days.
It is the most irregular one among all figures.
It may be illustrated that the user has a sudden trip and cannot connect to the OSN, or the work those days are busy so he pays little attention to OSN.
And those users may become reasonable constant users when he returns or finishes the work.
There is no effective strategy to crawl this kind of users, for we cannot predict their behavior, thus we cannot schedule the crawler well.
We treat those users as the inactive accounts to save crawling resource, and put them into the reasonable constant users once.
We found that they post many messages every day in the recent week.

\begin{figure}
  \centering
  \includegraphics[width=0.8\textwidth]{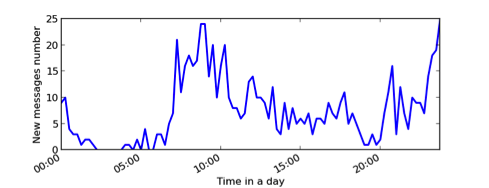}
  \caption{The Updating Ratio of an Reasonable Constant Account}
\end{figure}

3. Reasonable constant user.
Most valid accounts belong to this type.
For example, the users with work far from computer or mobile phone have to post messages after work and the users working with computer will post in work day in the office.
The frequency of the new messages is obviously influenced by the users' daily life, and the new messages often occur frequently in the afternoon and at night when the user takes a rest.
Hence we can predict their behavior by historic data.

Figure 3 shows the behavior of a reasonable constant user.
Such users love the OSN very much and post messages very regularly.
This kind of figures often has two peaks, the afternoon and the night.
The curve in the left of the peak grows up and the one after the peak goes down.

\begin{figure}
  \centering
  \includegraphics[width=0.8\textwidth]{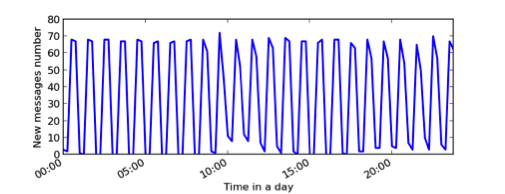}
  \caption{The Updating Ratio of an Authority Account}
\end{figure}

4. Authority account.
Such accounts are maintained by several clerks or just robots.
The content of those accounts is carefully updated far more frequent than ordinary people and is reviewed by more users.
For example, the New York Time updates news quickly on Twitter and has a large group of audience.

Figure 4 shows the behavior of an authority account.
The kind of figures is higher than the usual users' and has more peaks than the reasonable constant use.
From the result, there are one peak for each hour, and the peaks are almost the same high.
The reason is that clerks or robots may be asked to post messages once an hour.

\section{User Behavior Model And Crawl Strategy}

According to the behavior of the users, we build two models for effective crawling, the Poisson process model and Hash model.

The Poisson process model is built for inactive users, who behave similar to web pages, and the effective web crawling strategy such as SHARC \cite{denev2009sharc}, trade-offs \cite{baeza2008design} and sampling \cite{cho2002effective} can be applied to those users.

The Hash model records the principle of very active users' behavior.
For those users, we consider the frequency change in a day, and the change rate $\lambda_i$ is not same all the day long. Hence the Poisson process cannot describe it.
One effective way is to record the historic data change rate to predict the parameters for crawling.

\subsection{Poisson Process Model}
\label{sec:poi_model}
According to the observation of inactive users, we build a Poisson process model for their posting frequency, since those users behave like the updating of web pages, including changing constantly in a comparative long period (e.g. a month).
We assume that the change rate $\lambda_i$ can be estimated according to historic data, the number of audiences and channels, just as estimating change rate of pages by types, directory depths, and URL names.

However, the web crawling metrics require to be modified to fit OSN.
The SHARC \cite{denev2009sharc} define the blur for web pages, it describes the difference between the new and old pages of the same URL.
Yet this could not be applied to OSN.
The blur means the possibility that a page changes, but for the OSN, the question is the possible number of new messages, rather than whether a new message is post or not.
Thus the blur cannot describes the messages. To describe the freshness of messages in OSN, we define the potentiality for the users and the crawling target is to minimize the total potentiality.
We define the potentiality for OSN as follows.

\begin{mydef}
(Potentiality) Let $u_i$ be a OSN user crawled at time $t_i$, The potentiality of the user is the expected number of new messages between $t_i$ and query time $t$, averaged through observation interval [0, $n\Delta$]:
\end{mydef}

$$P({u_i},{t_i},n,\Delta ) = {1 \over {n\Delta }}\int \matrix{
  n\Delta  \hfill \cr
  0 \hfill \cr}  {\lambda _i}|t - {t_i}|dt = {{{\lambda _i}\omega ({t_i},n,\Delta )} \over {n\Delta }}$$,

where
$$\omega ({t_i},n,\Delta ) = t_i^2 - {t_i}n\Delta  + {{{{(n\Delta )}^2}} \over 2}$$

is the crawling penalty.

Let $U = (u_0,...,u_n)$ be OSN users crawled at the corresponding times $T = (t_0,...,t_n)$.
The blur of the crawling is the sum of the potentiality of the individual users is defined as,
$$P(U,T,n,\Delta ) = {1 \over {n\Delta }}\sum\limits_{i = 0}^n {{\lambda _i}\omega ({t_i},n,\Delta )} $$.

$¦Ø(t_i,n,\Delta)$ denotes the crawling penalty.
It depends on the crawling time $t_i$ and the period of the crawling interval $n\Delta$, but not on the user.
And we obtain Theorem~\ref{the:pro}.

\begin{mythe}
\label{the:pro}
(Properties of The Schedule Penalty) Double crawling delay will lead to fourfold crawling penalty and double potentiality:
\end{mythe}

$$\omega (i\Delta ,n,\Delta ) = {\Delta ^2}\omega (i,n,1)$$

$$P(U,T,n,\Delta ) = \Delta P(U,T,n,1)$$

For the interests of space, we omit the proof of Theorem~\ref{the:pro}.

For instance, the change rate for user $u_0$ to $u_5$ is $\lambda_0$=0, $\lambda_1$=1, $\lambda_2$=2, $\lambda_3$=3, $\lambda_4$=4 and $\lambda_5$=0, then we should crawl those users in the order $u_0$, $u_2$, $u_4$, $u_5$, $u_3$, $u_1$, thus we get the potentiality minimum.

\begin{codebox}
    \Procname{\proc{Algorithm 1 Crawl Schedule with Poisson Model}}
    \zi \verb'input': sorted users ($u_0,...,u_{n-1}$)
    \zi \verb'output': crawl schedule ($u_0^D,...,u_{n-1}^D$)

    \li \For $i \gets 0$ \To $\id{n}$
    \li \Do \If $i$ is even
            \li \Then $u_i^D = u_{i/2}$
            \li \ElseNoIf $u_i^D = u_{n-(i-1)/2}$
        \End
\end{codebox}

Algorithm 1 depicts the Process model algorithm for inactive users in OSN.
All the users are known advance.
We can sort and scan all the users only once to schedule the crawler.
The time and space complexity are both O($n$).
Thus we scan the user list only once.

\subsection{Hash Model}
\label{sec:hash}
According to the observation of reasonable constant users and authority accounts, we build the Hash model.
The changing rate $\lambda_i$ for those users is stable from the observations for a long time (e.g. a week or longer).
However, the new OSN messages are posted so frequently that they need to be crawled very frequently.
If we visit these users according to the averaged $\lambda_i$ of the day, we will waste much resource.

For example, we visit the user in the Figure 3 according to the Poisson process model.
In the Poisson process, the number of possible changes in each time unit is the same, so the time span between each extraction should be the same.
If we start crawling at 00:00 and crawl twice a day, it may be 00:00 and 12:00.
However, crawling at 03:00 and 19:00 seems the best strategy. Thus the Poisson process model does not fit those active users perfectly.

The number of active users' new messages changes frequently and is comparative randomly.
Hence it is hard to find a precise and suitable model.
On the other hand, the number of such kind of users is not large enough and a statistics model such as Gaussian model cannot be constructed according to the behaviors.
One effective way to predict the users' behavior is to record the historic data.

With such considerations, we define Hash model to obtain the messages of the active users better, including the reasonable constant users and authority accounts.
This model is built for the users who need to be crawled frequently, at least twice or more a day.
This model uses a hash table to record the number of new messages in a short recent period and the earlier data has less weight.
According to this historic data, we can calculate the possible number of new messages that are post in a given time span, so we can schedule the crawler better.

For example, we maintain an array of 24 bytes ($a_1,...,a_{24}$) to record the number of new messages posted by the same user in each hour.
$a_i$ values 0 at the beginning and the new value
$$a_i^{'}=a_i*0.5+n*0.5$$
where the parameter $n$ is the number of messages posted by the user at the ith hour of the day.
The weight for $k$ days before is $0.5^k$. As a result, the parameter for today is 0.5 and 0.25 for yesterday.

If a user does not post any messages for a few days, the values of the hash table decrease very fast.
To avoid such phenomenon, the longest crawling span threshold $s$ is required, that means, we will crawl the user at least once $s$ hours or days.

Although the hash-based method can be used for active user crawling, it is not suitable for some special cases.
For some special dates, such as the weekend, people behave differently from workday.
The experiment results of 100 users in 2 weeks show that 1496 messages were post on Tuesday while only 874 messages on Sunday.
Thus we could predict users' behavior by the last weekend data according to this feature.
Similar predications can be applied for the public holiday, such as the national day.
It is required to consider about the near and similar holiday data, or even the last year's vocation data with the hash model.

Algorithm 2 depicts the Hash model for one user.

\begin{codebox}
    \Procname{\proc{Algorithm 2 Crawl Schedule with Poisson Model}}
    \zi \verb'input': $a_1,...,a_k, n_1,...,n_k$
    \zi \verb'output': crawl time list $L$

    \li $lastCrawlTime = 0, sum_0=0-remainingMessage$
    \li \For $i \gets 1$ \To $\id{k}$
    \li \Do
            $a_i=a_i*0.5+n_i*0.5$
        \End
    \li $sum_1=a_1$
    \li \For $i \gets 2$ \To \id{k}
    \li \Do
            $sum_i = sum_{i-1}+a_i$
        \End

        \li \For $i \gets 1$ \To \id{k}
        \li \Do \If ($sum_i -sum_{lastCrawlTime} > c$) or ($i-lastCrawlTime>s$)
                        \li $L$.add($i$)
                        \li $lastCrawlTime = i$
            \End
            \li $remainingMessage = sum_k - sum_{lastCrawlTime}$
\end{codebox}

The length of the hash table is $k$ , we crawl $c$ messages each time and the user post $n_i$ messages yesterday at the time span $i$, $remainingMessage$ is the number of messages that are not crawled the day before, the $sum_i$ means the sum from $a_1$ to $a_i$ and $sum_0$ is the number of minus $remainingMessage$, thus number of the remaining messages that are not crawled yesterday will be count, and the longest crawling span threshold is $s$.
We input $a_1$,...,$a_k$, $n_1$,...,$n_k$ and it will output the crawl time list L.

First, we update the $lastCrawlTime$ and $sum_0$, and calculate the values of the hash table and the value of $sum_1$...$sum_n$.
Second, we scan the sums.
If there are enough messages to crawl ($sum_i - sum_{lastCrawltime} > c$) or the crawl time span exceed the threshold ($i - lastCrawltime > s$ ), then we add the time point $i$ to the crawl time list $L$ and update the $lastCrawltime$.

The time and space complexity is O($n$).

\section{Parallelization}
\label{sec:parallel}

As the information updates very fast on OSN, effective applications for OSN data require to collect messages on OSN at a high speed to keep the freshness of crawled data.
However, two factors prevent the high speed crawling.
One is the API quota for data accessing is seriously limited by the OSN platform (350 calls per hours on Twitter, and 150 calls on Sina Weibo).
Another is the computation required by information extraction from the crawled information and the size of social network is very large.
It is natural to design parallel algorithms to obtain fresh information from OSN.
With a number of IPs and machines, we are able to run the crawler on several computers at the same time to increase the throughout capability of the system.
For parallel processing, task scheduling is a crucial problem to solve.
For this problem, task scheduling is to assign crawling tasks for various users to machines to keep the balance of the loads of machines.

In this section, we will discuss the parallelization strategy for the crawler techniques presented before.
We propose load balance methods to make the crawler loads balanced on multiple machines.
In the following part of this section, first, we discuss parallel optimization of Poisson process model (in Section~\ref{sec:par_poisson}).
As for the Hash model, it focuses on scheduling crawling one user, rather than a large group of users as Poisson process model does.
It is not necessary to crawl one user with multiple machines, because the resource required to crawl one user is very few.
We can easily parallelize such process by using a few machines while every machine crawls different users.
Second, we propose two crawler system architectures (the centralized and distributed system), to make the crawler work in parallel (in Section~\ref{sec:arch}).
Both the Poisson process model and Hash model can be applied to the two system architecture.

\subsection{Parallel to Poisson Process Model}
\label{sec:par_poisson}

\begin{figure}
  \centering
  \includegraphics[width=0.35\textwidth]{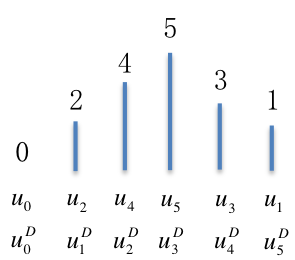}
  \caption{Organ Pipes in Poisson Process Model}
  \label{fig:organ}
\end{figure}

When the crawler development is based on Poisson process model as in Section ~\ref{sec:poi_model}, the figure of the message frequency and crawl sequence is like organ pipes as Figure ~\ref{fig:organ} shows.
The benefit of such model is the parameters are simple and the computation resource requirement for the parameters is small.

The hight in ~\ref{fig:organ} is the changing rate $\lambda$ in Poisson process.
For the OSN, the height means the user post frequency. By analyzing such model, we have following observations.

\begin{enumerate}
\item For those inactive and unimportant OSN users, we can collect their information at a low speed, e.g. once a month.

\item The $\lambda$, or post frequency, can be easily predicted according to the historic data of the target user.

\item Such a prediction does not cost much computing resource, what we need to do is to count how many messages the user post, and how long the time period is during his post behavior.
\end{enumerate}

Thus we can improve our crawler efficiency without calculating a lot.
The experiment shows it is 5.56\% to 12.14\% better than a Round-and-Robin style.

To make the best of the parallelization, we attempt to schedule the load for every machine as balanced as possible.

We propose two methods to achieve this goal, the Round-and-Robin method and the set-devision method.

\subsubsection{Round-and-Robin method}
\label{sec:rr}

A intuitive idea to allocate the tasks is to assign the crawl task of each user to the machines in Round-and-Robin style.

All crawl tasks $S$=\{$u_1,u_2,...,u_k$\}, each element in which is a crawl task for a user. The number of machines is denoted by $n$.
To make these computers work in parallel, we want to divide the sequence $S$ into $n$ parts, and then make each of the $n$ machines pick up one part and crawl for the users in this part. The Round-and-Robin method is to divide the sequence:

For machine $i$, we schedule machine it to crawl user $u_j(j \leq k)$ if $j$ mod $n = i$.

Thus we can get $n$ crawl lists, each of which is for one machine.

\begin{figure}
  \centering
  \includegraphics[width=0.7\textwidth]{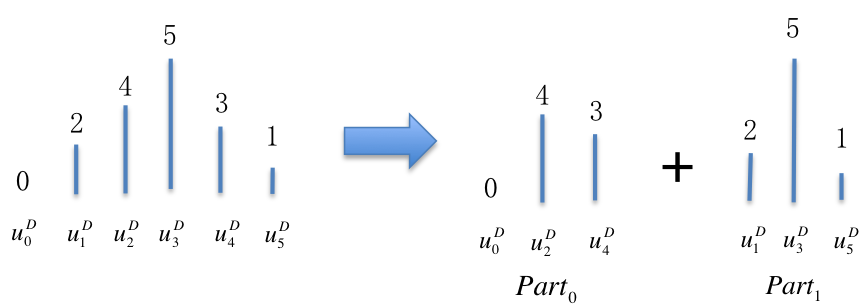}
  \caption{Round-and-Robin Method to Parallel Poisson Process Model into Two Parts}
  \label{fig:parallel_RR}
\end{figure}

Figure ~\ref{fig:parallel_RR} describes an example of the algorithm.
At the beginning, the to-crawl sequence of 6 users is $u_0^D,...u_5^D$, and we divide the users into two parts: $Part_0$ ( $u_0^D,u_2^D,u_4^D$) and $Part_1$ $u_1^D,u_3^D,u_5^D$.
For the user $u_i^D$ in $Part_0$, $i$ mod 2 is 0, and for the user $u_i^D$ in $Part_1$ $i$ mod 2 is 1.
We can conduct two machines, one to crawl $Part_0$ and another to $Part_1$.
Thus we can crawl the $u_0^D,...u_5^D$ in parallel.

Assume that in the users sequence sorted by their post frequency $u_0, u_1, ..., u_n$, the average post frequency difference between two adjoining users $f_{i+1}-f_i$ is a constant $\Delta$, where $f_i$ is the post frequency of user $u_i$.
In fact, the more users we count, the more stable $f_{i+1}-f_i$ is.
Then the workload difference between $Part_0$ and $Part_1$

\begin{eqnarray}
    \label{eq:part}
\sum Part_0 - \sum Part_1 = \left\{ \begin{array}{ll}
        0 & n \ \mbox{mod} \ 4=0 \\
f_0 & n \ \mbox{mod} \ 4=1 \\
-\Delta & n \ \mbox{mod} \ 4=2 \\
f_0-\Delta & n \ \mbox{mod} \ 4=3 \\
\end{array} \right.
\end{eqnarray}

\begin{proof}
We proof Equation~\ref{eq:part} in these four cases.
For the convenience of discussions, we further divide $Part_0$ and $Part_1$ into two segments respectively: $Part_{0A}$ and $Part_{0B}$, $Part_{1A}$ and $Part_{1B}$.

We use $f_i$ to represent the post frequency of user $u_i$, $f_i^D$ to the frequency of user $u_i^D$ in the Algorithm 1.
And there is
$$
f_i^D = \left\{ \begin{array}{ll}
        f_{i*2} & i < \frac{n+1}{2} \\
        f_{2(n-i)-1} & i \geq \frac{n+1}{2} \\
\end{array} \right.
$$

In the first case, $n$ mod 4 = 0.
$\lfloor \frac{n+1}{2}\rfloor$=$\frac{n}{2}$ and $\frac{n}{2}$ mod 2=0.
Thus, the workloads of the four parts are shown as follows.

\begin{eqnarray*}
    W_{Part_{0A}}&=&f_0^D + f_2^D + ... + f_{ \frac{n}{2}-2}^D \\
                 &= & \underbrace{ f_0+f_4+...+f_{n-4} }_{\mbox{$\frac{n}{4}-1$ items}} \\
    W_{Part_{1A}}&=&f_1^D + f_3^D + ... + f_{\frac{n}{2}-1}^D \\
                 &= & \underbrace{ f_2+f_6+...+f_{n-2} }_{\mbox{$\frac{n}{4}-1$ items}}\\
    W_{Part_{0B}} & = & f_{\frac{n}{2}}^D + f_{\frac{n}{2}+2}^D + ... + f_{n-2}^D \\
            & = & f_{2(n-\frac{n}{2})-1} + f_{2[n-(\frac{n}{2}+2)]-1} + ... + f_3 \\
            & = & \underbrace{ f_{n-1}+f_{n-5}+...+f_{3} }_{\mbox{$\frac{n}{4}-1$ items}}\\
    W_{Part_{1B}} &= &f_{\frac{n}{2}+1}^D + f_{\frac{n}{2}+3}^D + ... + f_{n-1}^D \\
            &=  &f_{2[n-(\frac{n}{2}+1)]-1} + f_{2[n-(\frac{n}{2}+3)]-1} + ... + f_1\\
            &=  & \underbrace{ f_{n-3}+f_{n-7}+...+f_{1} }_{\mbox{$\frac{n}{4}-1$ items}}\\
\end{eqnarray*}

Then the difference of the workloads of $Part_0$ and $Part_1$ is

\begin{eqnarray*}
    W_{Part_0} - W_{Part_1} & = & W_{Part_{0A}}+W_{Part_{0B}}-(W_{Part_{1A}}+W_{Part_{1B}}) \\
                            & = & [W_{Part_{0A}} - W_{Part_{1A}}] + [W_{Part_{0B}} + W_{Part_{1B}}] \\
                            & = & \underbrace{ [ (f_0 - f_2) + (f_4-f_6) + ... + (f_{n-4} - f_{n-2}) ]}_{\mbox{$\frac{n}{4}-1$ items}}\\
                            & & +  \underbrace{ [ (f_{n-1} - f_{n-3}) + (f_{n-5} - f_{n-7}) + ... + (f_3 - f_1)] }_{\mbox{$\frac{n}{4}-1$ items}}\\
                    & = & -(\frac{n}{4} -1)2\Delta + (\frac{n}{4} -1)2\Delta \\
                    & = & 0
\end{eqnarray*}

Since the $W_{Part_{0}}$ and $W_{Part_{1}}$ mean the $\sum Part_0$ and $\sum Part_1$ in Equation~\ref{eq:part}, when $n$ \ mod \ $4=0$, the $\sum Part_0 - \sum Part_1 = 0$ .

In the second case, $n$ mod 4 = 1.
$\lfloor \frac{n+1}{2}\rfloor$=$\frac{n+1}{2}$ and $\frac{n+1}{2}$ mod 2=0.
Thus, the workloads of the four parts are shown as follows.

\begin{eqnarray*}
    W_{Part_{0A}}&=&f_0^D + f_2^D + ... + f_{ \frac{n+1}{2}-1}^D \\
                 &= & \underbrace{f_0+f_4+...+f_{n-1} }_{\mbox{$\frac{n-1}{4}$ items}}\\
    W_{Part_{1A}}&=&f_1^D + f_3^D + ... + f_{\frac{n+1}{2}-2}^D \\
                 &= & \underbrace{ f_2+f_6+...+f_{n-3} }_{\mbox{$\frac{n-5}{4}$ items}}\\
    W_{Part_{0B}} & = & f_{\frac{n+1}{2}+1}^D + f_{\frac{n+1}{2}+3}^D + ... + f_{n-1}^D \\
                  & = & f_{2[n-(\frac{n+1}{2}+1)]-1} + f_{2[n-(\frac{n+1}{2}+3)]-1} + ... + f_{2[n-(n-1)]-1} \\
                  & = & \underbrace{ f_{n-4}+f_{n-8}+...+f_{1} }_{\mbox{$\frac{n-5}{4}$ items}}\\
    W_{Part_{1B}} &= &f_{\frac{n+1}{2}}^D + f_{\frac{n+1}{2}+2}^D + ... + f_{n-2}^D \\
                  &=  & f_{2(n-\frac{n+1}{2})-1} + f_{2[n-(\frac{n+1}{2}+2)]-1} + ... + f_3\\
                  &=  & \underbrace{ f_{n-2}+f_{n-6}+...+f_{3} }_{\mbox{$\frac{n-5}{4}$ items}}\\
\end{eqnarray*}

Then the difference of the workloads of $Part_0$ and $Part_1$ is

\begin{eqnarray*}
    W_{Part_0} - W_{Part_1} & = & W_{Part_{0A}}+W_{Part_{0B}}-(W_{Part_{1A}}+W_{Part_{1B}}) \\
                            & = & [W_{Part_{0A}} - W_{Part_{1A}}] + [W_{Part_{0B}} + W_{Part_{1B}}] \\
                            & = & [ f_0 + \underbrace{ (f_4 - f_2) + (f_8-f_6) + ... + (f_{n-1} - f_{n-3}) }_{\mbox{$\frac{n-5}{4} $ items}} ]\\
                            & & + [ \underbrace{(f_{n-4} - f_{n-2}) + (f_{n-8} - f_{n-6}) + ... + (f_1 - f_3)] }_{\mbox{$\frac{n-5}{4}$ items}}\\
                    & = & f_0 + \frac{n-5}{4}(2\Delta)- \frac{n-5}{4}(2\Delta) \\
                    & = & f_0
\end{eqnarray*}

Since the $W_{Part_{0}}$ and $W_{Part_{1}}$ mean the $\sum Part_0$ and $\sum Part_1$ in Equation~\ref{eq:part}, when $n$ \ mod \ $4=1$, the $\sum Part_0 - \sum Part_1 = f_0$ .

In the third case, $n$ mod 4 = 2.
$\lfloor \frac{n+1}{2}\rfloor$=$\frac{n}{2}$ and $\frac{n}{2}$ mod 2=1.
Thus, the workloads of the four parts are shown as follows.

\begin{eqnarray*}
    W_{Part_{0A}}&=&f_0^D + f_2^D + ... + f_{ \frac{n}{2}-1}^D \\
                 &= & \underbrace{f_0+f_4+...+f_{n-2} }_{\mbox{$\frac{n-2}{4}$ items}}\\
    W_{Part_{1A}}&=&f_1^D + f_3^D + ... + f_{\frac{n}{2}-2}^D \\
                 &= & \underbrace{ f_2+f_6+...+f_{n-4} }_{\mbox{$\frac{n-6}{4}$ items}}\\
    W_{Part_{0B}} & = & f_{\frac{n}{2}+1}^D + f_{\frac{n}{2}+3}^D + ... + f_{n-2}^D \\
                  & = & f_{2[n-(\frac{n}{2}+1)]-1} + f_{2[n-(\frac{n}{2}+3)]-1} + ... + f_3 \\
                  & = & \underbrace{ f_{n-3}+f_{n-7}+...+f_{3} }_{\mbox{$\frac{n-6}{4}$ items}}\\
    W_{Part_{1B}} &= &f_{\frac{n}{2}}^D + f_{\frac{n}{2}+2}^D + ... + f_{n-1}^D \\
                  &=  &f_{2(n-\frac{n}{2})-1} + f_{2[n-(\frac{n}{2}+2)]-1} + ... + f_1\\
                  &=  & \underbrace{f_{n-1}+f_{n-5}+...+f_{1} }_{\mbox{$\frac{n-2}{4}$ items}}\\
\end{eqnarray*}

Then the difference of the workloads of $Part_0$ and $Part_1$ is

\begin{eqnarray*}
    W_{Part_0} - W_{Part_1} & = & W_{Part_{0A}}+W_{Part_{0B}}-(W_{Part_{1A}}+W_{Part_{1B}}) \\
                            & = & [W_{Part_{0A}} - W_{Part_{1A}}] + [W_{Part_{0B}} + W_{Part_{1B}}] \\
                            & = & [ \underbrace{ (f_0 - f_2) + (f_4-f_6) + ... + (f_{n-6} - f_{n-4}) }_{\mbox{$\frac{n-6}{4}$ items}} + f_{n-2} ]\\
                            & & + [-f_{n-1} + \underbrace{ (f_{n-3} - f_{n-5}) + (f_{n-7} - f_{n-9}) + ... + (f_3 - f_5) }_{\mbox{$\frac{n-6}{4}$ items}}] \\
                            & = & [(\frac{n-6}{4})(-2\Delta) + f_{n-2} ] +[-f_{n-1} + (\frac{n-6}{4})(2\Delta) ]\\
                    & = & f_{n-2}-f_{n-1} \\
                    & = & -\Delta
\end{eqnarray*}

Since the $W_{Part_{0}}$ and $W_{Part_{1}}$ mean the $\sum Part_0$ and $\sum Part_1$ in Equation~\ref{eq:part}, when $n$ \ mod \ $4=2$, the $\sum Part_0 - \sum Part_1 = -\Delta$ .

In the forth case, $n$ mod 4 = 3.
$\lfloor \frac{n+1}{2}\rfloor$=$\frac{n+1}{2}$ and $\frac{n+1}{2}$ mod 2=0.
Thus, the workloads of the four parts are shown as follows.

\begin{eqnarray*}
    W_{Part_{0A}}&=&f_0^D + f_2^D + ... + f_{ \frac{n+1}{2}-2}^D \\
                 &= & \underbrace{f_0+f_4+...+f_{n-3} }_{\mbox{$\frac{n-3}{4}$ items}}\\
    W_{Part_{1A}}&=&f_1^D + f_3^D + ... + f_{\frac{n+1}{2}-1}^D \\
                 &= & \underbrace{ f_2+f_6+...+f_{n-1} }_{\mbox{$\frac{n-3}{4}$ items}}\\
    W_{Part_{0B}} & = & f_{\frac{n+1}{2}}^D + f_{\frac{n+1}{2}+2}^D + ... + f_{n-1}^D \\
                  & = & f_{2(n-\frac{n+1}{2})-1} + f_{2[n-(\frac{n+1}{2}+2)]-1} + ... + f_{2[n-(n-1)]-1} \\
                  & = & \underbrace{ f_{n-2}+f_{n-6}+...+f_{1} }_{\mbox{$\frac{n-3}{4}$ items}}\\
    W_{Part_{1B}} &= &f_{\frac{n+1}{2}+1}^D + f_{\frac{n+1}{2}+3}^D + ... + f_{n-2}^D \\
                  &=  &f_{2[n-(\frac{n+1}{2}+1)]-1} + f_{2[n-(\frac{n+1}{2}+3)]-1} + ... + f_3\\
                  &=  & \underbrace{f_{n-4}+f_{n-8}+...+f_{3} }_{\mbox{$\frac{n-7}{4}$ items}}\\
\end{eqnarray*}

Then the difference of the workloads of $Part_0$ and $Part_1$ is

\begin{eqnarray*}
    W_{Part_0} - W_{Part_1} & = & W_{Part_{0A}}+W_{Part_{0B}}-(W_{Part_{1A}}+W_{Part_{1B}}) \\
                    & = & [W_{Part_{0A}} - W_{Part_{1A}}] + [W_{Part_{0B}} + W_{Part_{1B}}] \\
                    & = & [  \underbrace{(f_0 - f_2) + (f_4-f_6) + ... + (f_{n-3} - f_{n-1}) }_{\mbox{$\frac{n-3}{4}$ items}}]\\
                    & & + [ \underbrace{(f_{n-2} - f_{n-4}) + (f_{n-6} - f_{n-8}) + ... + (f_5 - f_3)}_{\mbox{$\frac{n-7}{4}$ items}} + f_1] \\
                    & = & \frac{n-3}{4}(-2\Delta) + [(\frac{n-7}{4})(2\Delta) + f_1] \\
                    & = & f_1 - 2 \Delta \\
                    & = & f_0 - \Delta
\end{eqnarray*}

Since the $W_{Part_{0}}$ and $W_{Part_{1}}$ mean the $\sum Part_0$ and $\sum Part_1$ in Equation~\ref{eq:part}, when $n$ \ mod \ $4=3$, the $\sum Part_0 - \sum Part_1 = f_0 - \Delta$ .

\end{proof}

The workload difference between the two parts depends on the difference between the single user's post frequency $\Delta$ and the lowest post frequency $f_0$.

The Round-and-Robin algorithm has two advantages.
First, with $n$ increasing, the difference in workloads between two machines would not increase.
This property could assure the workload balance is still assured with large data size.
Second, the workload difference is negligible to a crawling computer.
In a real crawler, each machine crawls more than one thousand users a day, and the difference of one user is not influential.

The round-robin method only requires to scan the posting frequencies of users only once to partition each user $u_i$ into $Part_{i \mbox{\ mod\ } k}$.
Hence the space and time complexity for this algorithm is both O($n$), where $n$ is the number of users in the sequence.

\begin{figure}
  \centering
  \includegraphics[width=0.7\textwidth]{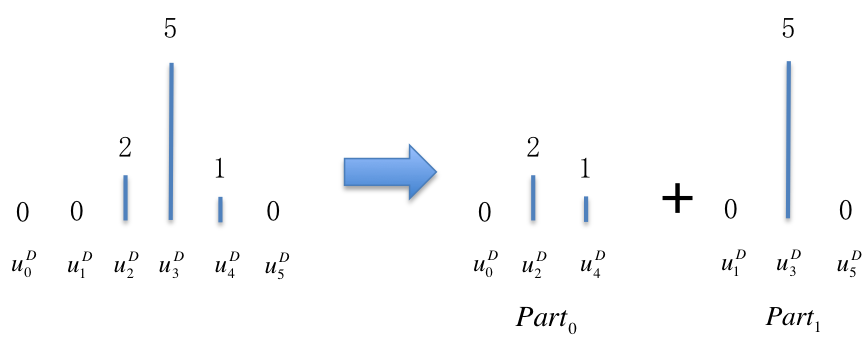}
  \caption{Round-and-Robin Method to Parallel Poisson Process Model Into Two Imbalanced Parts}
  \label{fig:parallel_diff}
\end{figure}

In some extreme cases, there are a few high active users, the load balance will be difficult.
Figure~\ref{fig:parallel_diff} shows such case. The user $u_3$ is very active while other users are quiet silent on OSN.
And the difference between $Part_0$ and $Part_1$ are larger than in Figure~\ref{fig:parallel_RR}.
Hence the load of two crawling sequences divided by the Round-and-Robin method is difficult to balance.
However, this is hardly met in the real crawling because of two reasons:

First, from the experimental results, the most active user among 88779 users post 2051 messages in a period of 4 years.
Comparing with the workload of a machine as millions of messages each day, Two thousand messages are not influential to the workload a crawling machine.
Therefore even the highest active users would hardly damage the load balance.
Second, in practise, the messages of much more users are to be crawled than the experiments. The more we collect, the more balance the Round-and-Robin method are.

\begin{figure}
  \centering
  \includegraphics[width=1\textwidth]{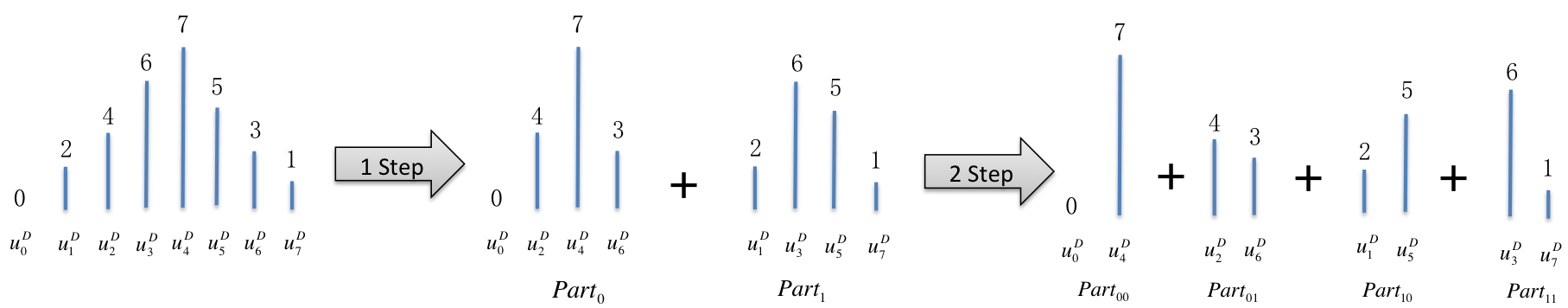}
  \caption{Round-and-Robin Method to Parallel Poisson Process Model into Four Parts}
  \label{fig:parallel_RR_multi}
\end{figure}

In cases that we need to divide the Round-and-Robin model into multiple parts, we can do the division recursively.
First, we divide the model into two parts, and then, we apply the same method to divide the two parts into four parts, and then eight or more parts.
Figure~\ref{fig:parallel_RR_multi} describes such a division.
At the beginning, there is just one part, and after Step 1, there are two parts, and after Step 2, there are four.
Thus we can crawl the original users with four machines at the same time.

This method can only divide the model into a power of 2 parts.
In face, since the number of machines is often a power of 2, the method is practical for such cases.

As for the time and space complexity, we proof that this algorithm is a PTAS as follow:

\begin{mythe}
The algorithm to parallel Poisson process model into multiple parts is a PTAS.
\end{mythe}

\begin{proof}
First, we attempt to prove that in the $i$th partition, the size of the largest share is at most $(1+\epsilon')^i\frac{n}{2^i}$ and the size of the smallest share is at most $(1-\epsilon')^i\frac{n}{2^i}$.
We prove this statement with induction.
As the basic step, for the first partition, according to the approximate ratio of FPTAS algorithm of SSP \cite{leiserson2001introduction}, the size of the smaller part should be larger than $(1-\epsilon')\frac{n}{2}$.
Accordingly, the size of the larger part is smaller than $n-(1-\epsilon')\frac{n}{2}$=$(1+\epsilon')\frac{n}{2}$.
The inductive assumption is that in the $(i-1)$th step, the size of the largest share is at most $(1+\epsilon')^{i-1}\frac{n}{2^{i-1}}$ and the size of the smallest share is at most $(1-\epsilon')^{i-1}\frac{n}{2^{i-1}}$.
According to the approximate ratio of FPTAS of SSP, the size of the smallest share of this step should be smaller than the lower bound of the smallest share, which is $(1-\epsilon')\cdot \frac{(1-\epsilon')^{i-1}\frac{n}{2^{i-1}}}{2}$=$(1-\epsilon')^i\frac{n}{2^i}$.
For the same reason, the size of the largest share of this step should be larger than the upper bound of the largest share, which is $(1+\epsilon')\cdot \frac{(1+\epsilon')^{i-1}\frac{n}{2^{i-1}}}{2}$=$(1+\epsilon')^i\frac{n}{2^i}$.
Then the statement is proven.

Then, we prove that the time complexity of the algorithm is in polynomial time of $n$, $k$.
All the algorithm has $\lceil \log k \rceil$ steps.
The time complexity of step $i$ is the polynomial time of $n$ and $k$, since the number of shares is $2^i\leq k$ and time complexity is in the polynomial time of $frac{n}{2^i}$ and $\frac{1}{\epsilon'}$.
Since there are at most $\lceil \log k \rceil$ steps.
The total time complexity is in polynomial time of $n$ and $k$.

Then we attempt to prove that the ratio of sizes of the largest share and the smallest share is smaller than a $1+\epsilon$ with a given $\epsilon$.
Since in the best case, the ratio of sizes of the largest share and the smallest share equals to 1.
Thus the ratio bound is $1+\epsilon$.

Since the size of the largest share is at most $(1+\epsilon')^i\frac{n}{2^i}$ and the size of the smallest share is at most $(1-\epsilon')^i\frac{n}{2^i}$ in step $i$, the ratio of the sizes of the largest share and the smallest share is at most $\frac{(1+\epsilon')^(\lceil \log k \rceil)}{(1-\epsilon')^(\lceil \log k \rceil)}$=$(\frac{1+\epsilon'}{1-\epsilon'})^(\lceil \log k \rceil)$.

Since $\epsilon'=\frac{R-1}{1+R}$, where $R=2^{\frac{\log(1+\epsilon)}{\lceil \log k \rceil}}$, we have $(1+R)\epsilon'=1-R$. Thus, $R=\frac{1+\epsilon'}{1-\epsilon'}$. That is,

$2^{\frac{\log(1+\epsilon)}{\lceil \log k \rceil}}$=$\frac{1+\epsilon'}{1-\epsilon'}$.

Then $\frac{\log(1+\epsilon)}{\lceil \log k \rceil}$=$\log \frac{1+\epsilon'}{1-\epsilon'}$.
We have $\log(1+\epsilon)$=$\lceil \log k \rceil \log \frac{1+\epsilon'}{1-\epsilon'}$.
Thus $1+\epsilon$=$2^{\log(1+\epsilon)}$=$2^{\lceil \log k \rceil \log \frac{1+\epsilon'}{1-\epsilon'}}$=$(\frac{1+\epsilon'}{1-\epsilon'})^(\lceil \log k \rceil)$.
It shows that the ratio of the sizes of the largest share and the smallest share is at most $1+\epsilon$.
Thus the ratio bound of the algorithm is $1+\epsilon$.
\end{proof}

\subsubsection{Set-Division method}

As discussed in the Section~\ref{sec:rr}, the round-robin strategy divides the previous part into several small parts, and each smaller parts are as the same large as possible.
The Set-Division method tries to solve the division by separating the previous part into a few number of smaller parts, and each smaller part contains different number of users.

If we treat the user post frequency as a set of integers, then the job schedule problem (JS for brief) is defined as follows.

Given a set of integers $S = {u_1,u_2,...,u_n}$, how to divide $S$ into $k$ sets $S_1, S_2, ..., S_k$ to make the difference between the $k$ sets

$$\sum \limits_{1 \le i < j \le k} {|\sum \limits_{i' \in {S_i}} {i'} }  - \sum\limits_{j' \in {S_j}} {j'} |$$
is the least.

We attempt to prove its hardness as Theorem~\ref{the:js}.

\begin{mythe}
\label{the:js}
JS is a NP-hard problem.
\end{mythe}

\begin{proof}
We attempt to reduce the Number Partitioning Problem (NPP)\cite{mertens2006easiest} to a special case of JS with $k$=2.

For a NPP problem with a given input of a set $S$=\{$a_1$, $a_2$, $\cdots$, $a_n$\} with integers, we construct a input of JS with the input set of integers as $S$ and $k=2$. The solution of such JS, $S_1$ and $S_2$, has minimal $|S_1-S_2|$. Thus this is a solution of such NPP problem. Since NPP is an NP-hard problem~\cite{gary1979computers} and the reduction can be accomplished in polynomial time clearly, JS is an NP-hard problem.
\end{proof}

The problem can be converted to the Subset Sum Problem (SSP). That is, given a set of integers $S = {u_1,u_2,...,u_n}$, how to select $m$ numbers from $S$ so that those $m$ numbers are close to but no more than a given constant $c$ \cite{soma2002exact}.

The SSP is NP-Hard \cite{gary1979computers}, but it can be solved in pseudo-polynomial time.
\cite{soma2002exact} gives such algorithm, and its time complexity is O(Max{($n-log_2c^2)c, c*log_2c$}), space complexity is O($n+c$).

\subsection{Parallel System Architectures}
In this subsection, we propose two parallel system architectures, a centralized and distributed ones.

In the centralized architecture, the clients are connected to a server, but all of them work independently.
While in the distributed architecture, all the computers work independently.
So you can apply either the Poisson process model or the Hash model to the two architectures, or apply the two models at the same time.

The major difference between the two architectures is whether the system has a central server.
For the centralized system, the server maintains the crawling sequences of all users, and assigns the crawling jobs to clients.
The clients are only responsible to do the crawling job sent by the server.
For the distributed system, every machine $i$ is in charge of a crawling task set $S_i$. They determine the assignment the task in $S_i$ to machines. Note for each task $u$ in $S_i$, machine $i$ just determines which machine will perform $u$ but $u$ does not have to be performed on machine $i$.

In this section, we will introduce these two architectures in Section and Section, respectively.



\subsubsection{Centralization Parallel System}
\label{sec:arch}

In the centralized crawler system, a center server is required to maintain the target crawling user sequence and schedule all clients' work.

The central server goes through the Poisson process model or Hash model and updates the model parameters.
It decides 1) who is the next user to crawl; and 2) which machine to crawl the user.
And other client machines all have a to-crawl user list.
Each client crawls information for the users in the user list one by one.

To make the best use of the API quota, when we want to collect a user's information, we first find the client machine whose to-do list is the shortest, and make it to do the job.
Hence the to-do list of each machine can be as the same long as possible, and the workload and the API quota of difference can be more similar.
If there is more than one machine with a shortest to-do list, then we randomly select a client to do the corresponding job.

In a word, the tasks for the server is:

1) Build the crawling model: Poisson process, Hash Model and so on.

2) Maintain the crawling sequence: decide who is the next user to crawl.

3) Send target user id to clients: let the client whose workload is currently minimal to crawl the target user.

4) Receive data from clients: store the data that clients collected.

And the tasks for a client is:

1) Crawl the target user: this is the job sent by the server.

2) Send the data to the server: so the server can manage the storage.

\begin{figure}
  \centering
  \includegraphics[width=0.7\textwidth]{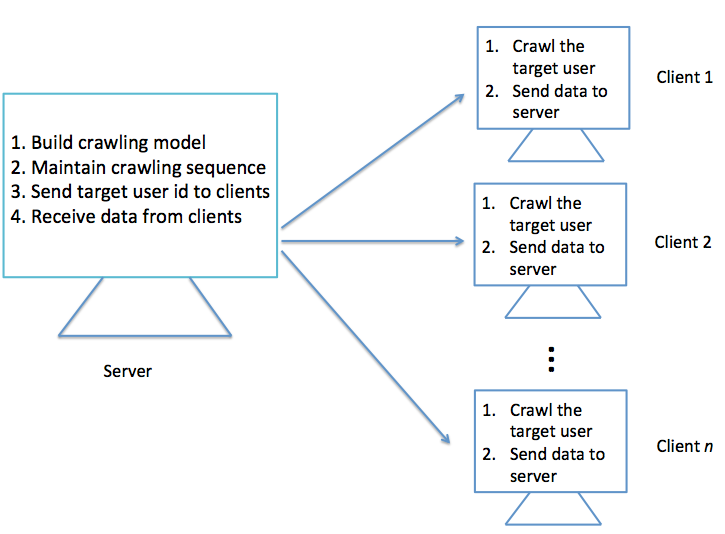}
  \caption{The Centralized Crawler Architecture}
  \label{fig:cen}
\end{figure}

Figure~\ref{fig:cen} describes such a centralized system.

It is convenient for a centralized system to do job scheduling, however, it may not make the best use of the bandwidth of the server.
As the task for the server is to compute and update the parameters for models, and that for the clients is to crawl, the limit for the server is computing resource while that for the client is API quota and bandwidth.
Therefore, it is possible that the server has run out of its computing resource while the clients still have enough API quota and bandwidth.
As a result, the centralized system may not make the best use of all machines.

On the other hand, the centralized system has a high requirement to the bandwidth between the server and clients.
If we use machines outside laboratory as clients, e.g. the plantlab, then it is hard to guarantee the bandwidth and the clients may lost connection.

To avoid the above two disadvantages, we propose a distributed architecture in the following subsection.

\subsection{Distributed Parallel System}

In a distributed system, there is no center server.
Every machine acts as both a server and client at the same time.

In the centralized architecture, the central server has a long list of OSN users that will be crawled.
The server is accessible to the users' data, and it manages the crawling strategy according to the users' behavior and crawling model. As a comparison, the client machines do not have any user list. The client machines are given the target users' names or ids, crawl according to the names or ids, and then return the users' data to the server.

In the distributed system, however, every machine has their own user list.
All of the machines manage their crawling dependently.
They are able to build different models, and rule the crawling strategy according to their own data respectively.
In a real crawling system, every machine runs two kinds of process: the Management Process and the Crawling Process.
The management process is responsible to the crawling strategy. It visits the users' data that has been collected before, runs crawling model (Poisson process model, Hash model or any other model), decides which user to crawl next, and sends the target user's name or id to the crawling process.
The crawling process is only visible to the target user's name or id sent by the management process.
The only duty of the crawling process is to crawl the target user, and then send the user's data back to the management process.
Because there are much more crawling job than management job, the number of crawling processes is much larger than that of management processes.
In other word, the management and crawling process serves like the server and client in the centralized system.

\begin{figure}
  \centering
  \includegraphics[width=0.8\textwidth]{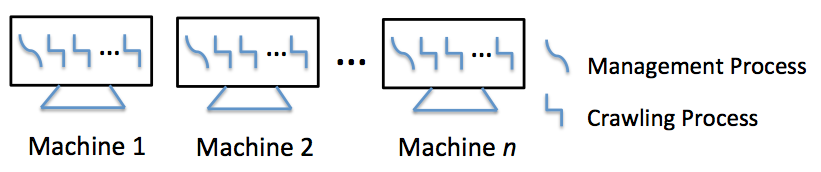}
  \caption{The Distributed Crawler Architecture}
  \label{fig:dis}
\end{figure}

Figure ~\ref{fig:dis} describes the distributed architecture of a crawling system.
The curves in the figure means the Management Process, while the fold lines means the Crawling Process.

Every machine under this architecture works independently, and they can use different crawling strategies of both Poisson process model or Hash model.
They also have different target user list.
They decide the crawling sequence of the users, and also do the crawling job assigned to it.

A number of processes are run on the computers at the same time.
There are two kinds of processes: 1) a computing process to maintain the model and adjust algorithm parameters, and 2) a few crawling processes to collect information from the online social network.
Because nowadays CPU always have multicores, to improve the performance of the crawling system, we can let the processes, no matter computing or crawling ones, run on different CPU cores.

As for the workload, in the centralized system, our purpose is to make the workload as balanced as possible as we discussed in Section~\ref{sec:rr}.
The centralized system is always built in the laboratory, because it requires a high speed network connection the server and clients.
In such cases, the client machines, or the clusters in the laboratory always have the same performance.
And to make the best use of the same machines, we just need to make the load as same as possible.

The distributed system, however, is always built outside the laboratory, for it has a lower requirement of the machines and network.
Under such circumstance, it is necessary to manage the workload according to the machine's performance.
In a distributed system, since the crawling work of every machine is independent, the workload is independent as well.
To make the workload between the machines balance, you can adjust by 1) switch the crawling algorithm, and 2) add or remove users from the user list of the computers.
For example, if you have a lot of inactive users to collect, and you want to crawl them just only once, then choose the Poisson process model, otherwise choose the Hash model.
And if a computer run out of its API quota, you can remove some users from its user list, and add the users to another computer whose API quota is enough.

However, the distributed architecture brings a problem: the data is stored distributively in many machines, which means we have to do more work to collect all the data in every machine.

In a word, the centralized system works better in laboratory, and the distributed system is more suitable outside the laboratory.

\section{Experimental Evaluation}

To verify the effectiveness and efficiency of our strategies, we perform experiments in this section.
The experiments are performed on a PC with 2.10 GHz Intel CPU and 4G RAM.
We crawl the Weibo.com, which is one of the hottest OSN in China.
The Weibo has 503 million registered users in 2012.12 and about 100 million messages are post everyday \cite{wiki-cite}.

We crawled the last 2,000 messages of 88,799 randomly selected users.
The result shows that 80,616,076 messages are collected.

\subsection{Experimental Results for Poisson Process Model}

To test the effectiveness of Poisson process model, we conducted the Poisson process model twice.

First, we crawled the 10,000 users in a period of 2 months from 1012.11.01 to 2013.01.01 , and accessed every user once to get the last 100 messages.
From observations, the Poisson process model crawled 421,722 messages, while the Round-and-Robin style crawled 376,053 messages.
Thus the Poisson process model is 12.14\% better.

Second, we crawled the 88,799 users in a period of 4 years from 2009.08.31 to 2013.09.01, and accessed every user once.
From observations, the Poisson process model crawled 7,369,498 messages, while the Round-and-Robin style crawled 7,147,965 messages.
Thus the Poisson process model is 3.10\% better.

3.10\% is lower than 12.14\%.
In fact, when the experimental period lasts longer and the total messages number grows larger, there will be more messages to collect, and every time we can access more messages.
And if we can collect more than 100 messages every time, the experimental result will be the same.
Therefore, the longer the experimental period, the more similar the results of various methods will be.

\subsection{Hash Model}

We crawl the 10,000 users by hash model.
We assume the crawling limit is 100 messages at one crawling, the longest crawling span threshold $s$ is 30 days, and the weight for the last one day (in the example, it is 0.5) ranges from 0.4 to 0.9 step by 0.1.
The result is shown in Table 1.
For comparisons, we also conduct the experiment for RR, and crawl one user 2 to 5 times during the experiment period.
The result is described in Table 2.

\begin{table}
\caption{The Hash Model Experiment Results}
\centering
\begin{tabular}{|c|c|c|c|}
        \hline
        Weight & Message Number & Total Crawl Time & Avg \#Msg. with \\ & & & 1 crawling \\
        \hline
        0.9 & 1451435 & 50421 & 28.79 \\
        0.8 & 1417908 & 44605 & 31.79 \\
        0.7 & 1368783 & 39446 & 34.70 \\
        0.6 & 1323012 & 35421 & 37.35 \\
        0.5 & 1255509 & 32211 & 38.98 \\
        0.4 & 1175464 & 29822 & 39.42 \\
        \hline
    \end{tabular}
\label{table:hash}
\end{table}


\begin{table}
\centering
    \caption{The RR Style Method Experiment Results}
    \begin{tabular}{|c|c|c|c|}
        \hline
        1 User Crawl Time & Total Crawl & Message Number & Avg. \#Msg. \\
                          & Time & & With 1 Crawling \\

        \hline
        5 & 50005 & 939964 & 18.80 \\
        4 & 40004 & 818039 & 20.45 \\
        3 & 30003 & 652641 & 21.75 \\
        2 & 20002 & 411086 & 20.55 \\
        \hline
    \end{tabular}
    \label{table:rr}
\end{table}

The data in Table 1 shows that with the weight increases, the total crawl time increases and more messages will be crawled, while the crawling efficiency decreases.
Thus the weight can be considered as a parameter to adjust the crawling efficiency and the limited crawling resource such as the API restriction.
The data in Table 2 shows that the crawling efficiency for the RR style method is reasonable stable when the crawling frequency is low.

\subsection{Parallelization}
We show the experimental result of the parallelization methods discussed in Section~\ref{sec:parallel} as follow:

\subsubsection{Parallel to Poisson Process Model}
First, we calculated the post frequencies (how many messages users post in a day) of 1000 to 7000 randomly selected user step by 1000, and tried both Round-and-Robin method and randomly select method to divide the frequencies into two parts.
Table~\ref{tab:PoiPro} shows the result.

From Table~\ref{tab:PoiPro}, we can find that the Round-and-Robin method is much better than the Random method.
In the worst experimental case (728.22 and 83.47), the workload difference of Round-and-Robin method is only 11.46\% of that of random method.

\begin{table}
\caption{The Round-and-Robin Parallelization Results 1}
\centering
\label{tab:PoiPro}
    \begin{tabular}{|c|c|c|c|}
        \hline
        User Num. & Fre. Tot. & Random Method Diff. & RR Diff. \\
        \hline
        10000 & 49943.17 & 1995.68 & 145.11 \\
        20000 & 96473.30 & 3413.03 & 37.54 \\
        30000 & 140484.31 & 5828.93 & 37.03 \\
        40000 & 187524.68 & 5421.92 & 85.72 \\
        50000 & 237834.48 & 728.22 & 83.47 \\
        60000 & 289173.84 & 8205.22 & 430.90 \\
        70000 & 335352.75 & 5651.32 & 343.31 \\
        \hline
    \end{tabular}
\end{table}

Second, we divide the 1000, 2000, 4000 and 8000 post frequencies into 16 parts recursively using both Round-and-Robin method and random method.
Table~\ref{tab:PoiProMulti} shows the result.

\begin{table}
\centering
\caption{The Round-and-Robin Parallelization Results 2}
\label{tab:PoiProMulti}
    \begin{tabular}{|c|c|c|c|}
        \hline
        User Num. & Fre. Tot. & Random Method Diff. & RR Diff. \\
        \hline
        10000 & 49943.17 & 2557.92 & 339.55 \\
        20000 & 96473.30 & 3096.89 & 276.63\\
        40000 & 187524.68 & 3264.10 & 185.39\\
        80000 & 381714.54 & 6585.32 & 643.81 \\
        \hline
    \end{tabular}
\end{table}

From Table~\ref{tab:PoiProMulti}, we can find that the Round-and-Robin method is much better than the Random method.
In the worst experimental case (2557.92 and 339.55), the workload difference of Round-and-Robin method is only 13.27\% of that of random method.

\subsubsection{Parallel to Centralization Architecture}
\label{sec:cen_res}

\begin{figure}
  \centering
  \includegraphics[width=0.6\textwidth]{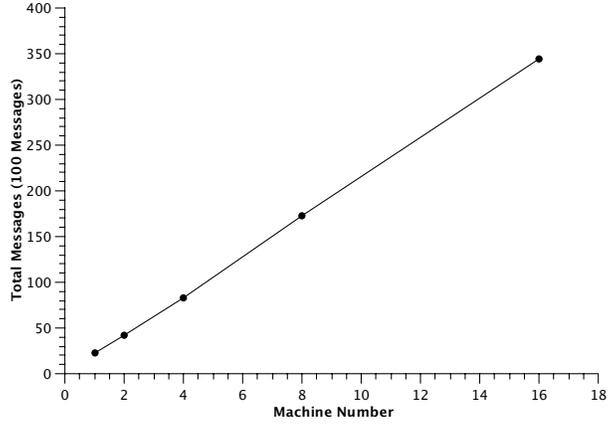}
  \caption{The Centralized Architecture Experimental Result}
  \label{fig:cen_res}
\end{figure}

To test the efficiency of the centralization architecture, we use 1, 2, 4, 8 and 16 machines respectively and crawl 2,000 users with one machine during a period of one year (from 2012.09.01 to 2013.09.1).
Table~\ref{tab:cen} and Figure~\ref{fig:cen_res} describe the experimental result.
In Table~\ref{tab:cen}, the 'Machine Num.' means the total number of machines in the experiment, the 'Tot. Messages' means the number of total messages that are crawled, and the 'Workload Diff.' means the difference between the maximum and minimum workload of the machines.
In the Figure~\ref{fig:cen_res}, the x-axis means the number of experimental machines, and the y-axis means the total messages crawled.

\begin{table}
\centering
\caption{The Centralized Architecture Experimental Result}
\label{tab:cen}
    \begin{tabular}{|c|c|c|}
        \hline
        Machine Num. & Tot. Messages & Workload Diff.\\
        \hline
        1 & 22474 & 0 \\
        2 & 42495 & 627 \\
        4 & 83271 & 4079 \\
        8 & 172473 & 3530\\
        16 & 344540 & 5008\\
        \hline
    \end{tabular}
\end{table}

From Table~\ref{tab:cen} and Figure~\ref{fig:cen_res}, we can find that the speed-up ratio is almost linear.

\subsubsection{Parallel to Distributed Architecture}

\begin{figure}
  \centering
  \includegraphics[width=0.6\textwidth]{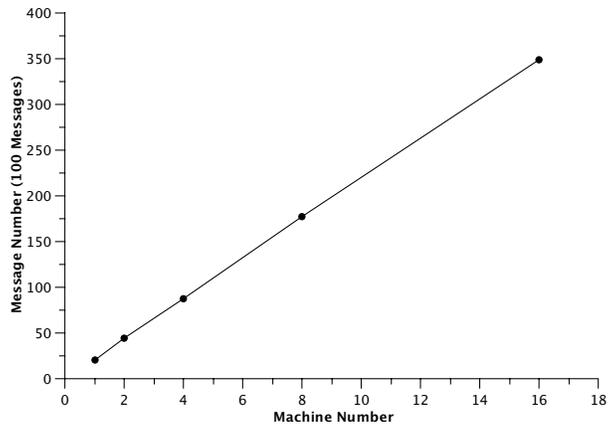}
  \caption{The Distributed Architecture Experimental Result}
  \label{fig:dis_res}
\end{figure}

To test the efficiency of the distributed architecture, we use 1, 2, 4, 8 and 16 machines respectively and crawl 2,000 users with one machine during a period of one year (from 2012.09.01 to 2013.09.1).
Table~\ref{tab:dis} and Figure~\ref{fig:dis_res} describe the experimental result.
The raws in Figure~\ref{tab:dis} and axises in Figure~\ref{fig:dis_res} have the same meaning in Figure~\ref{tab:dis} and Figure~\ref{fig:dis_res} mentioned in Section~\ref{sec:cen_res}

\begin{table}
\caption{The Distributed Architecture Experimental Result}
\centering
\label{tab:dis}
    \begin{tabular}{|c|c|c|}
        \hline
        Machine Num. & Tot. Messages & Workload Diff.\\
        \hline
        1 & 20916 & 0\\
        2 & 43886 & 1404\\
        4 & 87679 & 3575\\
        8 & 177534 & 4271\\
        16 & 348800 & 5269 \\
        \hline
    \end{tabular}
\end{table}

From the table, we can find that the speed-up ratio is almost linear.

Compare the distributed architecture with the centralized architecture, we find the workload difference of the distributed one is larger than the centralized one.
For the centralized system, the minimal slot to manage the workload is one crawling operation, while for the distributed system it's one target user.
We may crawl one user several times, so the slot for the centralized system is smaller and the workload is more balanced.

\section{Conclusion And Future Work}

To use the information in OSN effectively, it is necessary to obtain fresh OSN in-formation.
However, OSN crawler is quite different from web crawlers for the change rate is faster and the requirements for the latest messages are much more intensive.
Therefore, the traditional web crawl method cannot be applied for OSN information crawling.
To obtain fresh information in OSN with resource constraint, we classify users according to their behaviors.
Their behaviors are modeled and crawling algorithms are proposed according to the models.
Experimental results show that the Hash model is about 50\% better than the Round-and-Rabin method, and the Poisson process model is 12.14\% better than the RR method with randomly selected users.
And the parallelization method effectively controls the workload difference of the machines in the crawler system.
What's more, the centralized and distributed architectures all show a linear speed up with the number of machines.

There are some possible future research directions.
One is to crawl with different weight for different users, for the celebrities are more influential.
How to define the weights for users to crawl optimally remains a challenge.
Another direction is to obtain the hottest messages as early as possible.
Study of information transmit is required so that we can predict the hot pot in OSN.
As for the crawler system development, the centralized and distributed architectures may be tested with more machines and crawl more users, even the whole OSN users if possible.





\bibliographystyle{model1-num-names}
\bibliography{main}

\end{document}